\documentclass{article}

\usepackage{arxiv}

\usepackage[utf8]{inputenc} % allow utf-8 input
\usepackage[T1]{fontenc}    % use 8-bit T1 fonts
\usepackage{hyperref}       % hyperlinks
\usepackage{url}            % simple URL typesetting
\usepackage{booktabs}       % professional-quality tables
\usepackage{amsfonts}       % blackboard math symbols
\usepackage{nicefrac}       % compact symbols for 1/2, etc.
\usepackage{microtype}      % microtypography
\usepackage{amsmath}
\usepackage{bm}
\usepackage{graphicx}
\usepackage{amssymb}
\usepackage{color}

\title{Multilayer Network Analysis of the Drug Pipeline in the Global Pharmaceutical Industry}

\author{
  Hiromitsu Goto \\
  College of Science and Technology, Nihon University, Japan\\
    \texttt{hiromitsu.goto.phys@gmail.com} \\
   \And
  Wataru Souma\\
  College of Science and Technology, Nihon University, Japan\\
   \texttt{souma.wataru@nihon-u.ac.jp} \\
   \And
  Mari Jibu \\
  National Institute of Science and Technology Policy, Japan \\
   \texttt{mari.jibu@nistep.go.jp} \\
   \And
  Yuichi Ikeda \\
  Graduate school of Advanced Integrated Studies in Human Survivability, Kyoto University, Japan \\
   \texttt{ikeda.yuichi.2w@kyoto-u.ac.jp} \\
}
\begin{document}
\maketitle
%=========================================
\begin{abstract}
Generally, open innovation is a lucrative research topic within industries relying on innovation, such as the pharmaceutical industry, which are also known as knowledge-intensive industries. However, the dynamics of drug pipelines within a small-medium enterprise level in the global economy remains concerning.
To reveal the actual situation of pharmaceutical innovation, we investigate the feature of knowledge flows between the licensor and licensee in the drug pipeline based on a multilayer network constructed with the drug pipeline, global supply chain, and ownership data. 
Thus, our results demonstrate proven similarities between the knowledge flows in the drug pipeline among the supply chains, which generally agrees with the situation of pharmaceutical innovation collaborated with other industries, such as the artificial intelligence industry.

\end{abstract}
%=========================================

% keywords can be removed

%=========================================
\section{Introduction}
In the pharmaceutical industry, leading companies undergo shifts in the processes of research and development (R\&D),
production, and sales \cite{COMANOR2013106,SCHERER2010539,munos2009lessons}.
Particularly, the R\&D process is called the drug pipeline, which starts with a new drug discovery.
The development of new drugs requires pre-clinical testing, three stages of clinical trials, and approval in each country.
The shift of leading companies is recognized as an example of knowledge flows, which is an essential feature of academic collaboration \cite{Jibu201301,Jibu201302,Jibu201401,Jibu201402,Jibu201403,Jibu201404,Jibu201405,Jibu201406}.
However, the dynamics of the knowledge flows between the pharmaceutical companies in the drug pipeline remain a concern \cite{NARAYANA201418,MAZZOLA2015273}.

To understand the dynamics from a different perspective, we analyze the drug pipeline data of a pharmaceutical industry based on the multiplex network constructed with global supply-chain data, global ownership data, and various
financial statement data.
The multiplex networks are significant concepts for understanding complex systems, because many systems in the real world are constructed from various relations between components. 
For example, firms are connected by supply chain, ownership, a concurrent post of board members, and co-application of patents \cite{aoyama2010econophysics}.
Further examples of multiplex networks include transportation networks \cite{PhysRevLett.96.138701,ZOU20104406},
climatic systems \cite{Donges2011}, economic markets \cite{YANG20092435},
and energy-supply networks \cite{buldyrev2010catastrophic}.
The statistical mechanics in single networks are expanded to that in multiplex networks \cite{PhysRevE.87.062806}.

Therefore, our primary research goal is to show the emergence of propagation and the localization of the knowledge emerges by analyzing the drug pipeline data associated with global supply-chain data; hence, we consider three stage of analysis. 
First, we focus on the knowledge flows in an individual pipeline layer of the multiplex networks. 
The drug pipeline status falls into eight categories, ranging from discovery through launch, and there can be several current indications for a single drug candidate.
The shifts in R{\&}D and clinical trial processes depend on not only how much the drugs pipeline is developed but also the type of active indications the drug candidate possesses.
We need to focus on self-loops in the pipeline layer of the network, as they correspond to the pipeline of the original company.

Second, after the analysis of individual layers, we investigate the knowledge flows in the constructed multiplex network structure.
We confirm the overlaps of nodes and links to investigate the similarities between a pair of layers. Also, we identify the community structure of each network layer.

Finally, in the third stage, we discuss the superiority and uniqueness of the pharmaceutical industry in each country by considering the characteristics of the drug pipeline.

The rest of this paper is organized as follows. 
Section \ref{sec:data} describes the dataset used in our investigation. 
Section \ref{sec:representation} presents the knowledge flow of drug pipelines using a multilayer network (MLN) framework. 
Section \ref{sec:result} presents the analysis and results. 
Finally, Section \ref{sec:conclusion} concludes the paper.

%=========================================
\section{Data}
\label{sec:data}

We constructed a unique multiplex dataset that was beyond just a multiplex network superposed by a supply chain, ownership, and the drug pipeline layers. 
The drug pipeline data were acquired from Clarivate analytics. 
The Cortellis Competitive Intelligence provides pipeline data for various drug candidates in the R\&D stage and testing for approximately 7,000 pharmaceutical companies. 
We obtained the supply-chain and ownership data from the S\&P Capital IQ, which includes the supply-chain relation, ownership details, and companies' financial statements.

\paragraph{Drug Pipeline Data} 
We constructed a new database of drugs and the source and data procession as follows. 
We constructed the drug database, extracting all drug data from Clarivate Analytics and the Cortellis Competitive Intelligence database, on December 11, 2013. 
This database comprises several modules, among which we used drug and deal reports. 
Drug reports included the R\&D process to ensure the safety and efficiency of a drug: discovery, launched, clinical trial, pre-registration, and registered. 
The clinical trial can be categorized into three phases: Phase I (examinations for a drug safety carried out with healthy adults), Phase II (examinations for safety and effectiveness of a drug compared with existing drugs, carried out with a few patients), and Phase III (examinations for safety and effectiveness of a drug compared with existing drugs, carried out with many patients).
Deal reports include originator companies (i.e., licenser) and active companies (i.e., licensee). 
We indexed these companies' country and type. 
Regarding the company type, small- and medium-sized business firms are classified as ``private company," whereas large-sized businesses, universities, and government institutions are classified as ``public companies." 
The total number of drugs was 38,295, excluding those marking ``no development reported" in the highest status for 18 months before the extracted date.

\paragraph{Supply-chain and Ownership Data} 
The global supply-chain data and ownership data were constructed by collecting various company data from the website of the S\& P Capital IQ platform in 2018.
The Capital IQ dataset covers more than 500,000 companies with information on business relations in 217 countries in 159 industrial sectors defined by the S{\&}P, including all listed companies in the world.
The data include company ID, company name, country and location of company, company type, and primary industry as the node information. 
The supply-chain data also include examples of the business relation between the supplier and customer as the link information. 
Although various business relations that come under suppliers are supplier, creditor, franchiser, licenser, landlord, lessor, auditor, transfer agent, investor relations firm, and vendor, most of them are supplier and creditor. 
Here, the supplier indicates a company providing the products or services, while the creditor suggests a private, public, or institutional entity that makes funds available to others to borrow.
Furthermore, the ownership data include a list of shareholding companies and individuals for each company as the link information. The list comprises the top 100 companies and individuals with the ownership ratio data. 
Although the ownership ratio is supposed to be within 0–100 \%, a few exceptions are included.

Note that the collection days for the two datasets are different due to data availability. 
Therefore, this paper proposes an MLN using less than ideal, but actual, relations between companies to investigate the feature of knowledge flow in the drug pipeline.

%=========================================
\section{MLN Representation}\label{sec:representation}

In this section, we present the knowledge flow in drug pipelines using an MLN framework. 
In general, an MLN is a pair defined as $\mathcal{M}=(\mathcal{G},\mathcal{C})$, where $\mathcal{G}=\{G_{\alpha}; \alpha \in \{1,\cdots, M \} \}$ of the family of graphs $G_{\alpha}=(V_{\alpha},E_{\alpha})$ and $\mathcal{C}=\{E_{\alpha\beta}\subseteq V_{\alpha}\times V_{\beta}; \alpha,\beta \in \{1,\cdots, M \}, \alpha \neq \beta  \}$ is a set of interconnections between nodes of different layers $G_{\alpha}$ and $G_{\beta}$ with $\alpha\neq \beta$.
We use the drug pipeline, supply chain, and ownership data to define the MLN ($M=3$), which is composed of firms and institutions as nodes,
where the set of nodes of layer $G_{\alpha}$ is denoted as, $V_{\alpha} =\{v^{\alpha} _1, \cdots, v^{\alpha} _{N_{\alpha}}\}$.

The graphs $G_{\alpha}$ for each layer are defined as follows:
\begin{itemize}
    \item $G_{1}$: Knowledge flow network \\
    We describe the knowledge flow network as an unweighted directed graph, including self-loops and multiple edges. The drug pipeline data we collected from Clarivate analytics define edges from licenser $i$ to licensee $j$ with the status of R\&D process as edge attributions. 
    The launched drug pipeline data without transfer of license are represented as self-loops with ``launched" attribution, while the multiple edges represent those with different drugs.
    \item $G_{2}$: Supply-chain network\\
    The supply-chain network is an unweighted directed graph without multiple edges that represents the flows of products or services. The S\&P Capital IQ data define the flows between supplier $i$ and customer $j$. 
    \item $G_{3}$: Ownership network\\
    We define the edge from firm $i$ to $j$ when firm $j$ has a stake in firm $i$.
    Although the S\&P Capital IQ data include the ownership ratio, we define the ownership network as an unweighted directed graph to increase the number of duplicated nodes between layers. 
    Our ownership network represents the dependency flow, which is in the opposite direction to that typically used, because we assume that the firm knowledge tends to flow to these owners.
\end{itemize}
Note that we only extracted firms and ownership relations overlapping with different layers, and the number of owners of each firm was limited to 100 in the S\&P Capital IQ data. There was no interconnection between nodes of different layers in our MLN: $\mathcal{C}=\{\emptyset \}$. The primary purpose of our study is to investigate these interconnections between knowledge flows in drug pipelines and other layers.

The adjacency matrix of each layer $G_{\alpha}$ is denoted by $A^{[\alpha]}=\left(a_{ij} ^{\alpha}\right)$, where the element $a_{ij} ^{\alpha}$ corresponds to the number of edges from node $i$ to $j$ in the $\alpha$-th layer.
The in- and out-degrees of a node $i$ of the MLN are defined as vectors:
\begin{equation}
    \bm{k}_{\text{in},i} = \left(k^{[1]} _{\text{in},i}, ~~k^{[2]} _{\text{in},i}, ~~k^{[3]} _{\text{in},i}\right)~~~\text{and}~~~~\bm{k}_{\text{out},i} = \left(k^{[1]} _{\text{out},i}, ~~k^{[2]} _{\text{out},i}, ~~k^{[3]} _{\text{out},i}\right),
\end{equation}
where $k^{[\alpha]} _{\text{in},i}$ and  $k^{[\alpha]} _{\text{out},i}$ are the in- and out-degrees of node $i$ in the $\alpha$-th layer, i.e., $k^{[\alpha]} _{\text{in},i} =\sum_j a_{ji} ^{\alpha}$ and $k^{[\alpha]} _{\text{out},i} = \sum_j a_{ij} ^{\alpha}$.
Similarly, we denote the number of self-loops of node $i$ in the $\alpha$-th layer as, $\ell^{[\alpha]} _i= a_{ii} ^{\alpha}$.
In this paper, it is important to recognize the flows as self-loops because self-loops in the knowledge flow network $G_1$ correspond to the accumulation of knowledge.
Thus, we count the number of flows in each layer by ``edges" between different two nodes and ``self-loops": the number of edges of node $i$ from/to different nodes is denoted as
\begin{equation}
    m^{[\alpha]} _{\text{in},i}=k^{[\alpha]} _{\text{in},i} - \ell^{[\alpha]} _i ~~~~~~\text{and}~~~~~~~m^{[\alpha]} _{\text{out},i}=k^{[\alpha]} _{\text{out},i} - \ell^{[\alpha]} _i~,
\end{equation}
and their total number of them is
\begin{equation}
    M_{\alpha}=\frac{1}{2}\sum_{i,j=1,i\neq j} ^{N_{\alpha}} a_{ij} ^{\alpha}~~~~~~\text{and}~~~~~~~L_{\alpha}=\sum_i^{N_{\alpha}} a_{ii} ^{\alpha}~.
\end{equation}
Table~\ref{tab:stat_net}  summarizes the numbers of nodes, edges, and self-loops at each layer. 
We list the numbers of successfully combined Coritellis and Capital IQ datasets. 
The first layer for the knowledge flows based on the drug pipeline data of Coritellis is mostly composed of educational institutions and pharmaceutical firms, including ventures.
Only a small part of the nodes of the supply-chain and ownership networks is combined with those of the knowledge flow layer because the second and third layers are based on the Capital IQ datasets covering the whole industrial sectors. 
Since the discovered drug pipelines tend to appear as self-loops, the knowledge flows based on them contain several self-loops. 
Our framework has the potential to clarify the reality of closed innovation in the pharmaceutical industry in terms of the whole economic system.

%+++++++++++++++++++++++++++++++++
\begin{table}[ht]
\caption{Numbers of nodes, edges, and self-loops before combining Coritellis and Capital IQ datasets. We also list the number after the combination.}
\vspace{2mm}
    \centering
    \begin{tabular}{lrrrrrr}
    \toprule
    %\hline
    {} & \multicolumn{2}{c}{ $G_1$: Knowledge Flow} & \multicolumn{2}{c}{ $G_2$: Supply Chain}  & \multicolumn{2}{c}{ $G_3$: Ownership}  \\
    {} & Whole & (Combined) & Whole & (Combined) &  Whole & (Combined) \\
    \midrule
    %\hline
    $N_{\alpha}$: \# of nodes      &          7,785 &    (2,379) &      503,840 &    (1,954) &     69,552 &      (830) \\
    $M_{\alpha}$: \# of edges      &         11,478 &    (3,769) &    1,379,344 &    (3,115) &  1,147,937 &      (158) \\
    $L_{\alpha}$: \# of self-loops &         20,876 &   (10,927) &            5 &        (0) &         24 &        (0) \\
    \bottomrule
    %\hline
    \end{tabular}
    \label{tab:stat_net}
\end{table}
%+++++++++++++++++++++++++++++++++

%=================
\subsection*{The construction of an MLN for knowledge flows based on drug pipelines} 

\begin{figure}
    \centering
    \includegraphics[width=4.5in]{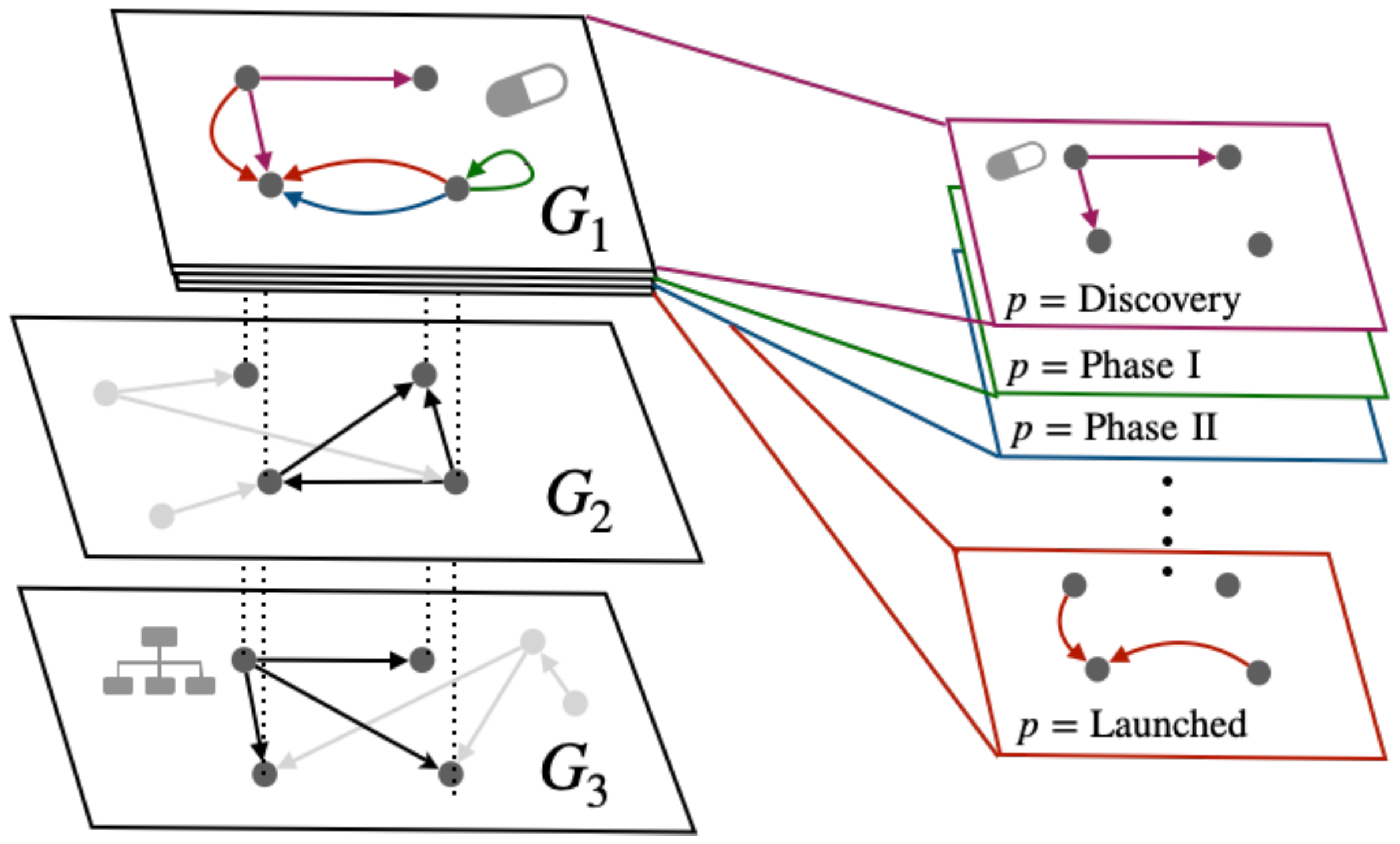}
    \caption{Overview of MLN representation. We use drug pipeline, supply-chain, and ownership data to define the MLN ($M=3$), which is composed of firms and institutions as nodes. Then, we construct MLN for knowledge flow based on the edge attribution about the status of drug pipeline.}
    \label{fig:multiNet}
\end{figure}

To understand the dynamics of drug pipelines in terms of networks, 
we propose a framework for the knowledge flow of drug pipelines $G_1=(V_1,E_1)$ decomposed using edge attributions, as $E_1 = \bigcup _{p}E_1 ^{p}$, where $E_1 ^p$ represents the edge in the knowledge flow network at the status $p\in\{$Discovery, Phase I-III Clinical, Pre-registration, Registered, Launched$\}$.
In response to extension, we add the status layer index $p$ to the definitions of the first layer characteristics such as in-degree of $i$-th node $k^{[1,p]} _{\text{in},i}$ and the total number of edges $M^{p}_1$.
Figure~\ref{fig:multiNet} illustrates the multiplex representation we take, and Table~\ref{tab:stat_pipeline} shows the number of edges and self-loops for each pipeline status. 
Note that the number of drug pipelines at the launched status is cumulative. 
The self-loops at the discovery status are dominant since they correspond to drug discovery seeds and are candidates of the future Phase I clinical status. 
However, the self-loop components for all statuses are dominant in the drug pipeline, except for the launched status, indicating that the pharmaceutical firms tend to develop drug pipelines from discovery to launch. 
This representation can help us understand when and how the knowledge of drug pipelines flows.

%+++++++++++++++++++++++++++++++++
%+++++++++++++++++++++++++++++++++

%+++++++++++++++++++++++++++++++++
\begin{table}[ht]
    \caption{Numbers of edges and self-loops for each pipeline status in the knowledge flow layer.}
    \vspace{3mm}
    \centering
    \begin{tabular}{lrrr}
    \toprule
    %\hline
    $p$: pipeline status &  $M_1 ^p$: \# of edges &  $L_1 ^p$: \# of self-loops & ~~~~~~Total \\
    \midrule
    %\hline
    Discovery          &  4,488 &     13,668 &  18,156 \\
    Phase I Clinical   &    818 &      1,716 &   2,534 \\
    Phase II Clinical  &  1,255 &      1,941 &   3,196 \\
    Phase III Clinical &    538 &        630 &   1,168 \\
    Pre-registration   &    165 &        189 &     354 \\
    Registered         &    120 &        155 &     275 \\
    Launched           &  4,065 &      2,527 &   6,592 \\
    \bottomrule
    %\hline
    \end{tabular}
    \label{tab:stat_pipeline}
\end{table}
%+++++++++++++++++++++++++++++++++

%=========================================
\section{Analysis and Results}\label{sec:result}

%=================
\subsection*{Characteristics of knowledge flow in drug pipelines} 
Here we show the characteristics of the drug pipelines using the knowledge flow network representation. 
The drug pipeline begins with the discovery of a new drug candidate, which requires pre-clinical testing, three stages of clinical trials, and approval to launch. 
Generally, the venture firms or educational institutions provide grants to the discovery of drug seeds and the licensee; for example, leading firms advance development to launch, incurring considerable R\&D expenses, and these features vary from country to country.

\begin{figure}
    \centering
    \includegraphics[width=6in]{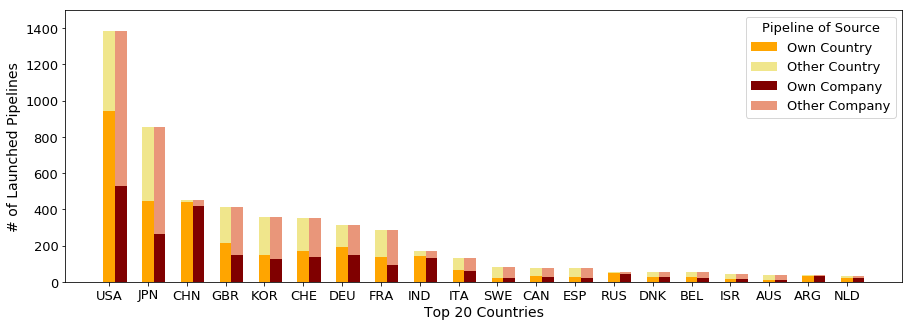}
    \caption{Number of drug pipelines at the launched status by countries. We show top 20 countries with the self‐sufficiency rates by own country and own company.}
    \label{fig:launch_hist_own}
\end{figure}

Figure~\ref{fig:launch_hist_own} shows the number of drug pipelines at the launched status by country, where the numbers correspond to the sum of in-degrees of the licensee at the launched status by individual countries, i.e., $\sum_{i\in a}k^{[1,p]} _{\text{in},i}$ at $p=\{$Launched$\}$ by country $a$.
Moreover, we distinguish the drug pipelines at the launched status based on the self-sufficiency of individual countries and firms, compared to the licenser of drug pipelines. 
It is known that USA has launched several drug pipelines. 
We found that most of the top 20 countries have discovered and launched more than half of their drug pipelines in their own countries, and the drug pipelines in China (CHN) tend to be developed in her country compared to different other countries.

%+++++++++++++++++++++++++++++++++
%+++++++++++++++++++++++++++++++++

\begin{figure}[htbp]
 \begin{minipage}{1\hsize}
  \begin{center}
   \includegraphics[width=5.5in]{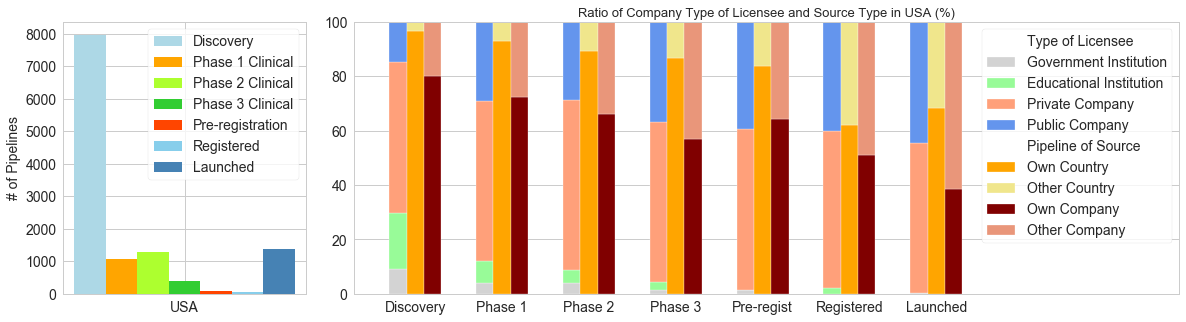}
   \\
   (a) USA
    \vspace{2mm}
  \end{center}
 \end{minipage}
 \begin{minipage}{1\hsize}
 \begin{center}
  \includegraphics[width=5.5in]{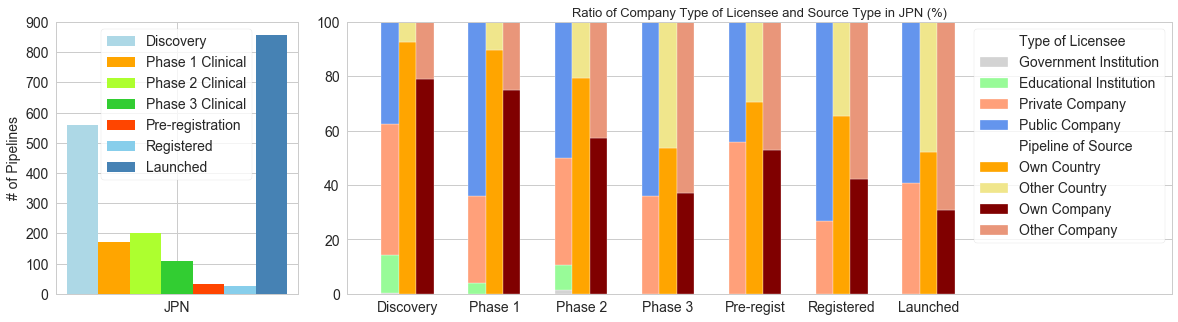}
   \\
   (b) JPN
    \vspace{2mm}
 \end{center}
 \end{minipage}
 \begin{minipage}{1\hsize}
  \begin{center}
   \includegraphics[width=5.5in]{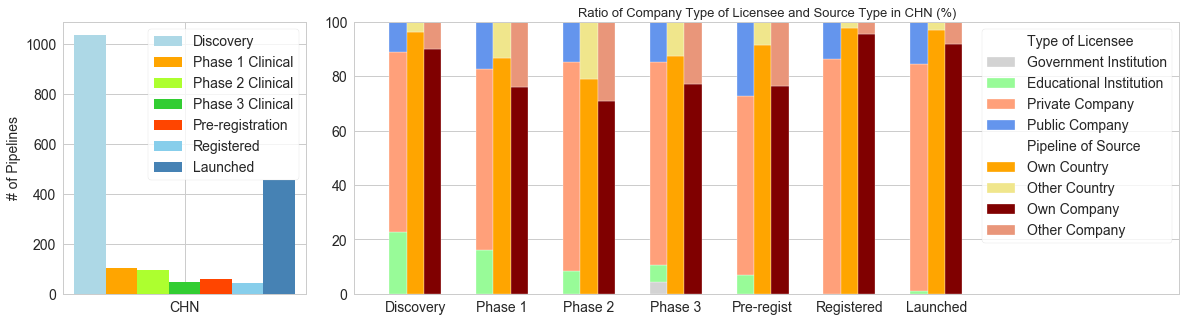}
   \\
   (c) CHN
    \vspace{2mm}
  \end{center}
 \end{minipage}
 \begin{minipage}{1\hsize}
 \begin{center}
  \includegraphics[width=5.5in]{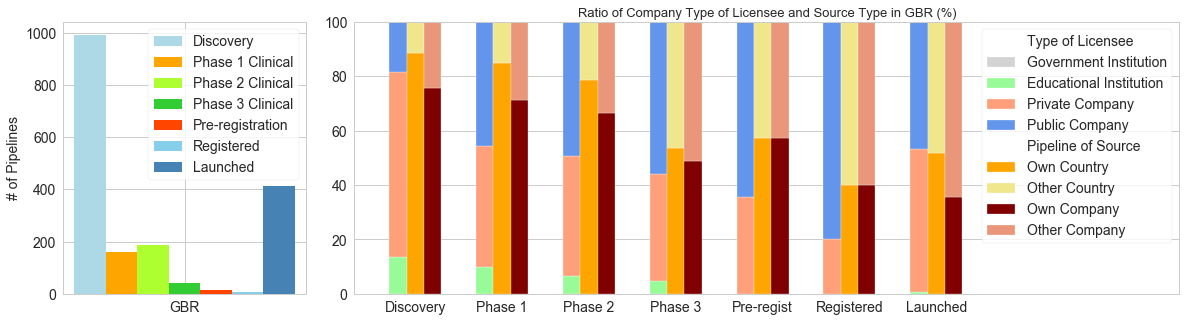}
   \\
   (d) GBR
    \vspace{2mm}
 \end{center}
 \end{minipage}
 \caption{Number of drug pipelines for each status, and these fraction of type of licenser and licensee. We show top 4 countries having drug pipelines at the launched status, (a) USA, (b) JPN, (c) CHN and (d) GBR.}
 \label{fig:pipeline_status}
\end{figure}

The licensee of drug pipeline is expected to change, owing to the considerable R\&D expense.
To characterize the changes from the viewpoint of type of business entity, we add the node attribution to categorize the type of licenser or licensee: $V_1 = \bigcup _{a}V_1 ^{a}$ where $a\in\{$Government Institution, Educational Institution, Private Company, Public Company$\}$. 
Figure~\ref{fig:pipeline_status} shows the number of drug pipelines for each status for the top 4 countries having drug pipelines at the launched status, the ratio of components in terms of business entity, and self-sufficiency.
As shown in Figure~\ref{fig:pipeline_status} (a), USA has around 8,000 drug discovery seeds, which is more than eight times compared to that possessed by the other countries. 
As expected, the government and educational institutions tend to release the license at an early stage of the R\&D process and the license acquisition rate of the public company; for example, as the number of leading firms increases as the process reaches the launched status. 
Moreover, it is only in Japan (JPN) that the number of drug pipelines at the discovery status is smaller than that at the launched status. 
As the lack of drug discovery seeds nowadays will decline the pharmaceutical industry of the country in the future, countries such as JPN need to obtain drug pipelines from other countries.

%+++++++++++++++++++++++++++++++++
%+++++++++++++++++++++++++++++++++

%+++++++++++++++++++++++++++++++++
\begin{figure}[thbp]
 \begin{minipage}{0.49\hsize}
  \begin{center}
   \includegraphics[width=2.5in]{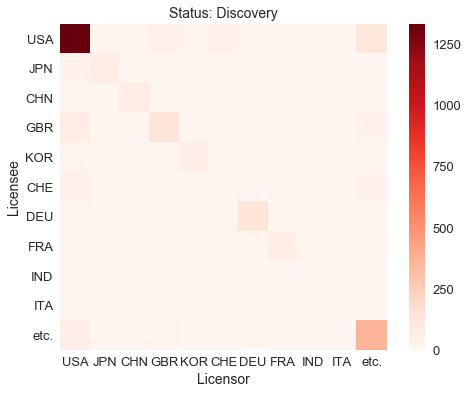}
  \end{center}
 \end{minipage}
 \begin{minipage}{0.49\hsize}
 \begin{center}
  \includegraphics[width=2.5in]{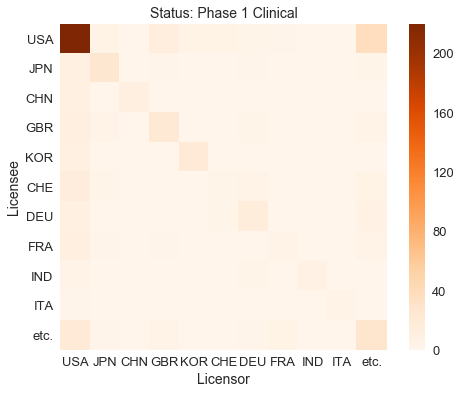}
 \end{center}
 \end{minipage}
 \begin{minipage}{0.49\hsize}
  \begin{center}
   \includegraphics[width=2.5in]{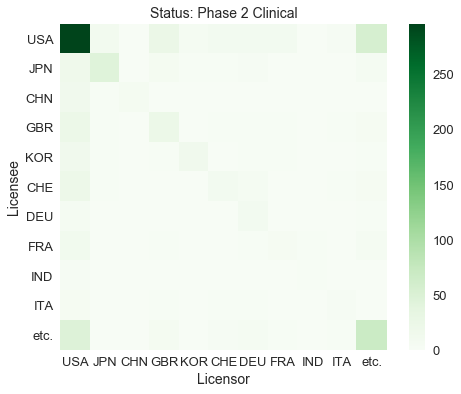}
  \end{center}
 \end{minipage}
 \begin{minipage}{0.49\hsize}
 \begin{center}
  \includegraphics[width=2.5in]{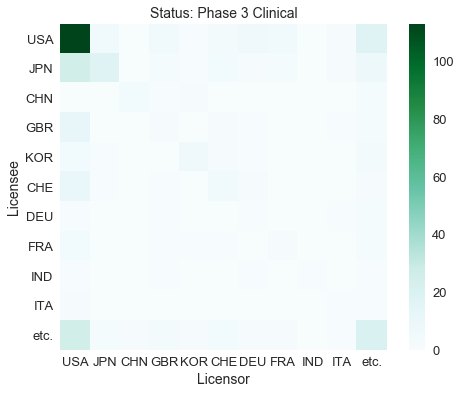}
 \end{center}
 \end{minipage}
 \begin{minipage}{0.49\hsize}
  \begin{center}
   \includegraphics[width=2.5in]{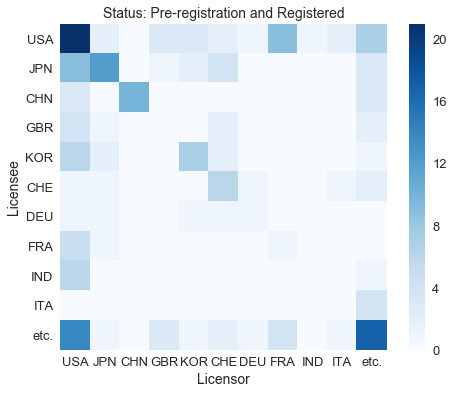}
  \end{center}
 \end{minipage}
 \begin{minipage}{0.49\hsize}
 \begin{center}
  \includegraphics[width=2.5in]{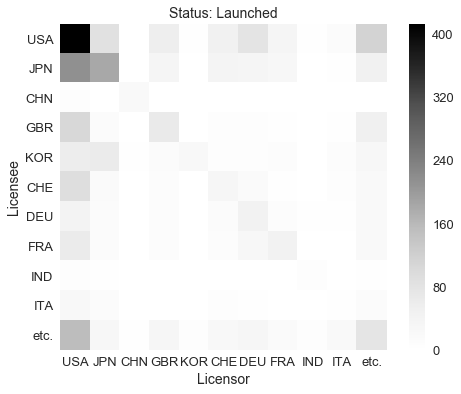}
 \end{center}
 \end{minipage}
 \caption{Knowledge flows of drug pipelines between licenser and licensee at the country level per pipeline status, focused on the top 10 countries having drug pipelines at the launched status. Each element represents the number of drug pipelines between countries with the self-loops contributions being ignored.}
    \label{fig:edge_status}
\end{figure}
%+++++++++++++++++++++++++++++++++

Finally, Figure~\ref{fig:edge_status} shows the knowledge flows in the drug pipelines between the licenser and licensee at the country level per pipeline status. 
Here we focus on the top 10 countries having drug pipelines at the launched status, and aggregate the pre-registration and registered statuses, because of the availability of a few pipelines, and ignore the self-loops, in order to investigate the flows between different business entities. 
Thus, on the whole, the knowledge of drug pipelines flows in the country and the firms in USA provide the drug pipelines to the other countries. 
Moreover, the knowledge of drug pipelines seems to flow not only in the direction from firms in USA to the others but also in the opposite direction.

%=================================
\subsection*{Single-layer Analysis}
The primary purpose of this study is to show the emergence of propagation and the localization of knowledge, by analyzing not only drug pipeline data but also supply-chain and ownership data. 
As discussed previously, the knowledge of drug pipelines is localized in individual companies and countries. 
They expect to appear as self-loops and the circular flows, such as strongly connected components (SCCs), in the network. 
Moreover, it is highly important to discuss knowledge circulation in terms of open innovation, because drug pipeline development requires significant R\&D expenses. 
Before discussing the multilayer analysis, therefore, we investigate the macroscopic structure of isolated networks.

\begin{figure}
    \centering
    \includegraphics[width=5in]{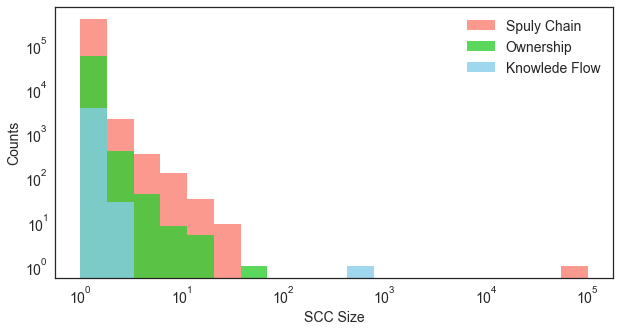}
    \caption{Distribution of strongly connected component (SCC) size for each layer. The size of SCC is measured as the number of nodes.}
    \label{fig:scc_size}
\end{figure}

%=================
\paragraph{Bow-tie Structure} 
In general, the giant weakly connected components (GWCCs) of a directed network can be decomposed as giant strongly connected components (GSCCs), which is the largest size of SCC in the GWCC, and its upstream and downstream portions (i.e., IN and OUT), known as the bow-tie decomposition in the Web \cite{broder2000graph}.
This decomposition can help us understand the hierarchical and circular flows of the networks from a macroscopic perspective. First, we compute how many and how large are the networks for each layer containing the SCC.
As shown in Figure~\ref{fig:scc_size}, the knowledge flow network of the drug pipeline is immensely large in terms of GSCC, compared to the other SCC, which is the same as the supply-chain network.

\begin{figure}[thbp]
 \begin{minipage}{0.33\hsize}
  \begin{center}
   \includegraphics[width=2in]{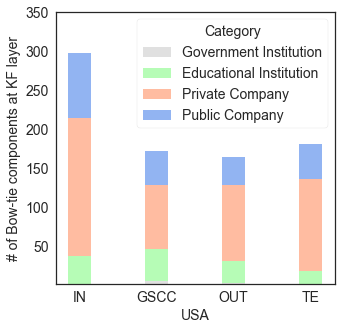}
  \end{center}
 \end{minipage}
 \begin{minipage}{0.33\hsize}
 \begin{center}
  \includegraphics[width=2in]{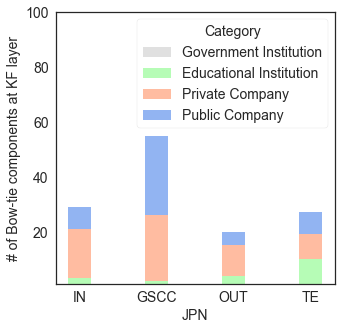}
 \end{center}
 \end{minipage}
 \begin{minipage}{0.33\hsize}
  \begin{center}
   \includegraphics[width=2in]{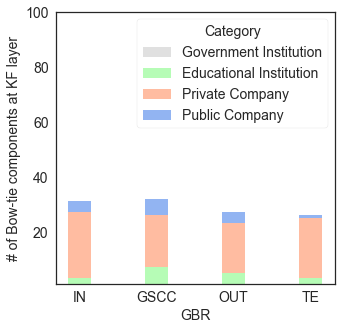}
  \end{center}
 \end{minipage}
 \caption{Number of bow-tie components in the knowledge flow network by countries. We categorized each component using the type of business entities and only showed the top 3 countries located in the GWCC.}
 \label{fig:bowtie_category}
\end{figure}

Furthermore, we show the results of bow-tie decomposition for each layer network in Table~\ref{tab:bowtie}, where we define the GWCC components not belonging to the GSCC, IN, and OUT components as Tendril (TE). 
Clearly, the ownership network has quite different bow-tie structure structures than the others. 
The ownership network does not have a large SCC, such as a large circular flow of control.
Conversely, the bow-tie structure of the knowledge flow layer has similar fractions of components to those of the supply-chain layer. 
This suggests that the knowledge flows of drug pipelines have the same macroscopic features as those of the supply-chain network. 
Moreover, as discussed previously, it is necessary to investigate the role of the countries in knowledge circulation. 
As shown in Figure~\ref{fig:bowtie_category}, the largest fraction of firms in USA are located on the upstream, and several public companies in JPN belong to GSCC of the knowledge flow networks. 
Although the GWCC of the knowledge flow network is mainly composed of the firms in USA, it is assumed that these firms provide the drug pipeline seeds to the whole system and that the leading firm in JPN contributes to the circular flow of knowledge.

%+++++++++++++++++++++++++++++++++
%+++++++++++++++++++++++++++++++++

%+++++++++++++++++++++++++++++++++
\begin{table}[thbp]
    \caption{The sizes of different components based on bow-tie decomposition for each layer.}
    \vspace{3mm}
    \centering
    \begin{tabular}{lrrrrrr}
     %\hline
     \toprule
    {} & \multicolumn{2}{c}{ Knowledge Flow} & \multicolumn{2}{c}{ Supply Chain}  & \multicolumn{2}{c}{ Ownership}  \\
    %\midrule
    Componets & Counts & Ratio (\%) &   Counts & Ratio (\%) &   Counts & Ratio (\%) \\
    
     \midrule
    IN   &  1,020 &     23.88 &  100,976 &     21.17 & 33,011 &     56.82  \\ 
    GSCC &    498 &     11.66 &   80,225 &     16.82 &      63 &      0.11 \\
    OUT  &  1,313 &     30.74 &  218,344 &     45.79 &     501 &      0.86 \\
    TE   &  1,440 &     33.72 &   77,332 &     16.22 &  24,525 &     42.21 \\
     \midrule
    GWCC &  4,271 &     100.00&  476,877 &    100.00 &  58,100 &   100.00  \\
    \bottomrule
    \end{tabular}
    \label{tab:bowtie}
\end{table}
%+++++++++++++++++++++++++++++++++

%+++++++++++++++++++++++++++++++++
%+++++++++++++++++++++++++++++++++

%=================
\paragraph{Community Structure} 
Community detection is a powerful tool for explaining the structural properties of densely connected networks. 
To find communities in GWCC of the layers, we use here the map equation method \cite{rosvall2008maps}, 
popularly known as ``Infomap," which is one of the best performing community detection method~\cite{lancichinetti2009community}.
The map equation method is a flow-based and information-theoretic approach, whose objective is to find an efficient code for minimizing the length of the description of the random walk for generating a module partition $\mathcal{M}$ dividing $n$ nodes into $m$ communities. Then, the average single-step description length is defined as
\begin{equation}
    L(\mathcal{M})=q_{\curvearrowleft}H(\mathcal{Q})+\sum^{m}_{i=1}p_{i\circlearrowright}H(\mathcal{P}_i)~.
\end{equation}
The first term arises from the movements of the random walker across modules, where $q_{\curvearrowleft}$ is the probability that the random walker switches communities, and $H(\mathcal{Q})$ depicts the average description length of the community index codewords given by the Shannon entropy.
The second term arises from the intra-community movement of random walker, where the weight $p_{i\circlearrowright}$ represents the fraction of the movements within the community, and $H(\mathcal{P}_i)$ represents the entropy of the intra-community movement.
Moreover, this method has been extended to a hierarchical map equation \cite{rosvall2011multilevel}, which decomposes a network into communities, and sub-communities.

Thus, we employ the hierarchical map equation method to reveal the communities in each layer.
Table ~\ref{tab:com_sum} summarizes the modular-level statistics. In general,  a large part of the nodes belongs to 2nd-level communities, where we count the number of nodes in irreducible communities, and the results are found to be similar to those of the study of the Japanese production network \cite{chakraborty2018hierarchical}.
To compare the community structure for each layer from a macroscopic perspective, here we limit our discussion of the properties of communities up to the 2nd level, and characterize each community by checking the majority attributes of nodes in the community.

First, in Table~\ref{tab:com_KF}, we show the characteristics of the largest community at the 1st layer for the knowledge flow layer and the three largest communities at the 2nd level, where we use the attributions of country, bow-tie component, and category of company for this layer and measure the community size by using the number of nodes in the community. 
Then, comparing the fraction of attributions at the 1st and 2nd levels, each community is found to have similar characteristics, including 10\% of public companies. 
The knowledge of drug pipelines flows in the set of several private companies and a few leading firms, and is widely distributed in the largest community.

Second, we demonstrate the characteristics of the communities in the other layers, where we use attributes of the primary industry instead of the category of companies. 
As shown in Table~\ref{tab:com_SC}, the primary industry of the firms forms the communities at the 1st level in the supply-chain layer, and the 3rd largest community labeled 4th has a significant fraction of the pharmaceutical sector. To investigate the relation between the knowledge flows in the pharmaceutical industry, we investigate the 2nd-level sub-communities in the community labeled 4th in Table~\ref{tab:com_SC_2}. 
Note that the community comprising the pharmaceutical firms in CHN is detected as the 2nd largest sub-community at the supply-chain layer. 
These trends are in line with the previously discussed structure of the knowledge flows of drug pipelines in CHN, which is extremely close.

Finally, Table~\ref{tab:com_OW} shows the characteristics of the six largest communities at the 1st level. Table~\ref{tab:com_OW_2} shows the features of the three largest sub-communities of the largest community at the 1st level in the ownership layer. 
Consequently, there are no significant characteristics in the largest communities, based on ownership. 
Although it is difficult to relate the knowledge flows of drug pipelines in this stage, future work should focus on the relation to the community of dependency flow, in contrast with the knowledge flows by the mergers and acquisitions (M\&A).

%+++++++++++++++++++++++++++++++++
\begin{table}[thbp]
\caption{Modular-level statistics for the communities detected by using the multi-coding Infomap method. "\#com." is the number of all communities, "\#irr.com." is the number of irreducible communities, which are communities that do not have any sub-communities. "\#nodes" represents the number of nodes such as firms and institutions in irreducible communities.}
    \vspace{2mm}
    \centering
    \begin{tabular}{rrrrrrrrrr}
    \toprule
    %\hline
    {} & \multicolumn{3}{c}{$G_1$: Knowledge Flow} & \multicolumn{3}{c}{$G_2$: Supply Chain}  & \multicolumn{3}{c}{$G_3$: Ownership}  \\
    Level & \#com.  & \#irr.com.  & \#nodes & \#com. & \#irr.com. & \#nodes  & \#com.  & \#irr.com.  & \#nodes \\
    \midrule
    %\hline
    1 &            784 &                 778 &             778 &            425 &                  70 &             479 &            447 &                 429 &             709 \\
2 &            410 &                 320 &           2,754 &         41,532 &              35,586 &         381,343 &          7,819 &               6,689 &          29,953 \\
3 &            287 &                 279 &             701 &         22,591 &              22,155 &          92,599 &         12,185 &              11,671 &          23,706 \\
4 &             16 &                  16 &              38 &          1,028 &               1,022 &           2,425 &          1,731 &               1,687 &           3,531 \\
5 &            {} &                 {} &             {} &             15 &                  15 &              31 &            112 &                 112 &             201 \\
    \bottomrule
    %\hline
    \end{tabular}
    \label{tab:com_sum}
\end{table}
%+++++++++++++++++++++++++++++++++

%+++++++++++++++++++++++++++++++++
\begin{table}[thbp]
    \caption{Characteristics of the largest community at the 1st level for the knowledge flow layer and the three largest communities at the 2nd level. The figures in brackets indicate the fraction of attribution, and we have only listed the attributions having a share of more than 5\%.}
    \vspace{3mm}
    \centering
    \begin{tabular}{lllll}
    \toprule
    %\hline
    \multicolumn{5}{l}{$G_1$: Knowledge Flow}  \\
    Index &  Size & Country & Category &  Bowtie \\
    %\hline
    \midrule
    1:    &  3,380 & 
    \begin{tabular}{l}
    USA (48.5)\\JPN (7.6)\\GBR (7.2)
    \end{tabular}  &  
    \begin{tabular}{l}
    Private Company (76.7)\\Public Company (12.6)\\Educational Institution (9.9)
    \end{tabular} &
    \begin{tabular}{l}
    TE (31.5)\\IN (27.3)\\OUT (27.1)\\GSCC (14.1)
    \end{tabular}
    \\
    \bottomrule
     \\
    \\
    \multicolumn{5}{l}{3 largest sub-communities}  \\
    \midrule
    1:1:    &  68 & 
    \begin{tabular}{l}
    USA (66.7)\\ISR (7.4)
    \end{tabular}  &  
    \begin{tabular}{l}
    Private Company (73.5)\\Educational Institution (11.8)\\Public Company (10.3)
    \end{tabular} &
    \begin{tabular}{l}
    OUT (36.8)\\IN (32.4)\\GSCC (17.6)\\TE (13.2)
    \end{tabular}
    \\
    \midrule
    1:3:    &  66 & 
    \begin{tabular}{l}
    USA (59.4)\\CAN (6.2)\\JPN (6.2)\\GBR (6.2)
    \end{tabular}  &  
    \begin{tabular}{l}
    Private Company (83.3)\\Public Company (10.6)
    \end{tabular} &
    \begin{tabular}{l}
    OUT (37.9)\\GSCC (25.8)\\TE (18.2)\\IN (18.2)
    \end{tabular}
    \\
    \midrule
    1:7:    &  63 & 
    \begin{tabular}{l}
    USA (50.0)\\CAN (7.7)\\GBR (7.7)
    \end{tabular}  &  
    \begin{tabular}{l}
    Private Company (81.0)\\Public Company (12.7)\\Educational Institution (6.3)
    \end{tabular} &
    \begin{tabular}{l}
    OUT (42.9)\\IN (28.6)\\TE (15.9)\\GSCC (12.7)
    \end{tabular}
    \\
    \bottomrule
    \end{tabular}
    \label{tab:com_KF}
\end{table}
%+++++++++++++++++++++++++++++++++

%+++++++++++++++++++++++++++++++++
\begin{table}[thbp]
    \caption{Characteristics of the six largest communities at the 1st level for the supply-chain layer. The figures in brackets indicate the fraction of attribution, and we have only listed the attributions having a share of above 5\%.}
    \vspace{3mm}
    \centering
    \begin{tabular}{lllll}
    \toprule
    \multicolumn{5}{l}{$G_2$: Supply Chain}  \\
    Index &  Size & Country & Primary Industry &  Bowtie \\
    \midrule
    3:    &  43,636 & 
    \begin{tabular}{l}
    USA (32.3)\\CHN (9.1)\\JPN (7.5)\\TWN (6.3)
    \end{tabular}  &  
    \begin{tabular}{l}
    Application Software (7.6)\\Technology Distributors (7.3)\\Semiconductors (5.2)
    \end{tabular} &
    \begin{tabular}{l}
    OUT (53.5)\\IN (17.3)\\GSCC (15.1)\\TE (14.1)
    \end{tabular}
    \\
    \midrule
    1:    &  30,972 & 
    \begin{tabular}{l}
    USA (43.2)\\CHN (9.4)\\GBR (7.1)
    \end{tabular}  &  
    \begin{tabular}{l}
    Application Software (8.8)\\Movies and Entertainment (7.5)
    \end{tabular} &
    \begin{tabular}{l}
    OUT (44.5)\\IN (22.9)\\GSCC (20.2)\\TE (12.5)
    \end{tabular}
    \\
    \midrule
    4:    &  30,430 & 
    \begin{tabular}{l}
    USA (52.0)\\CHN (6.0)\\GBR (5.2)
    \end{tabular}  &  
    \begin{tabular}{l}
    Pharmaceuticals (15.7)\\Biotechnology (11.1)\\Health Care Equipment (8.2)\\Health Care Facilities (7.0)\\Health Care Distributors (5.2)
    \end{tabular} &
    \begin{tabular}{l}
    OUT (35.6)\\IN (28.0)\\GSCC (19.7)\\TE (16.7)
    \end{tabular}
    \\
    \midrule
    5:    &  28,298 & 
    \begin{tabular}{l}
    IND (70.4)\\USA (6.1)
    \end{tabular}  &  
    \begin{tabular}{l}
    Construction and Engineering (5.3)
    \end{tabular} &
    \begin{tabular}{l}
    OUT (48.6)\\GSCC (23.8)\\IN (18.5)\\TE (9.0)
    \end{tabular}
    \\
    \midrule
    2:    &  25,365 & 
    \begin{tabular}{l}
    USA (47.6)\\GBR (5.8)\\CHN (5.4)
    \end{tabular}  &  
    \begin{tabular}{l}
    Apparel, Accessories and Luxury Goods (8.5)\\Packaged Foods and Meats (7.1)
    \end{tabular} &
    \begin{tabular}{l}
    OUT (41.5)\\IN (23.7)\\GSCC (17.7)\\TE (17.1)
    \end{tabular}
    \\
    \midrule
    6:    &  17,857 & 
    \begin{tabular}{l}
    USA (36.2)\\CHN (8.3)\\GBR (5.6)
    \end{tabular}  &  
    \begin{tabular}{l}
    Oil and Gas Exploration and Production (15.1)\\Oil and Gas Storage and Transportation (7.8)\\Oil and Gas Equipment and Services (6.7)
    \end{tabular} &
    \begin{tabular}{l}
    OUT (36.8)\\IN (27.7)\\GSCC (23.2)\\TE (12.4)
    \end{tabular}
    \\
    \bottomrule
    \end{tabular}
%    \begin{itemize}
%    \color{red}{
%    \item The percentages for Primary Industry was wrong. I corrected them.}
%\end{itemize}
    \label{tab:com_SC}
\end{table}
%+++++++++++++++++++++++++++++++++

%+++++++++++++++++++++++++++++++++
\begin{table}[thbp]
        \caption{Characteristics of the three largest communities in the 4th community at the 2nd level for the supply-chain layer. The figures in brackets indicate the fraction of attribution, and we have only listed the attributions with a share above 5\%.}
    \vspace{3mm}
    \centering
    \begin{tabular}{lllll}
    \toprule
    \multicolumn{5}{l}{$G_2$: Supply Chain}  \\
    Index &  Size & Country & Primary Industry &  Bowtie \\
    \midrule
    4:2:    &  792 & 
    \begin{tabular}{l}
    USA (97.0)
    \end{tabular}  &  
    \begin{tabular}{l}
    Construction and Engineering (70.5)
    \end{tabular} &
    \begin{tabular}{l}
    IN (94.3)
    \end{tabular}
    \\
    \midrule
    4:7:    &  509 & 
    \begin{tabular}{l}
    CHN (99.0)
    \end{tabular}  &  
    \begin{tabular}{l}
    Health Care Distributors (47.5)\\Pharmaceuticals (35.6)\\Biotechnology (7.6)
    \end{tabular} &
    \begin{tabular}{l}
    OUT (41.8)\\GSCC (31.2)\\IN (26.9)
    \end{tabular}
    \\
    \midrule
    4:22:    &  364 & 
    \begin{tabular}{l}
    USA (47.0)\\DEU (7.5)\\GBR (6.9)
    \end{tabular}  &  
    \begin{tabular}{l}
    Biotechnology (52.4)\\Life Sciences Tools and Services (17.7)\\Health Care Equipment (8.1)\\Health Care Distributors (6.5)\\Pharmaceuticals (5.6)
    \end{tabular} &
    \begin{tabular}{l}
    TE (84.6)\\OUT (14.6)
    \end{tabular}
    \\
    \bottomrule
    \end{tabular}
    \label{tab:com_SC_2}
\end{table}
%+++++++++++++++++++++++++++++++++

%+++++++++++++++++++++++++++++++++
\begin{table}[thbp]
    \caption{Characteristics of the six largest communities at the 1st level for the ownership layer. The figures in brackets indicate the fraction of attribution, and we have only listed the attributions with a share above 5\%.}
    \vspace{3mm}
    \centering
    \begin{tabular}{lllll}
    \toprule
    \multicolumn{5}{l}{$G_3$: Ownership}  \\
    Index &  Size & Country & Primary Industry &  Bowtie \\
    \midrule
    1:    & 54845 & 
    \begin{tabular}{l}
    USA (21.9)\\IND (9.0)\\CAN (6.4)\\CHN (5.7)\\AUS (5.2)\\GBR (5.2)\\JPN (5.1)
    \end{tabular}  &  
    \begin{tabular}{l}
    Asset Management and Custody Banks (10.1)
    \end{tabular} &
    \begin{tabular}{l}
    IN (56.8)\\TE (42.2)
    \end{tabular}
    \\
    \midrule
    7:    &  644 & 
    \begin{tabular}{l}
    USA (24.0)\\IND (10.3)\\CAN (6.2)\\AUS (5.9)\\JPN (5.4)
    \end{tabular}  &  
    \begin{tabular}{l}
    Asset Management and Custody Banks (11.6)
    \end{tabular} &
    \begin{tabular}{l}
    IN (57.3)\\TE (41.1)
    \end{tabular}
    \\
    \midrule
    6:    &  532 & 
    \begin{tabular}{l}
    USA (22.2)\\IND (8.3)\\CAN (7.5)\\AUS (6.2)\\JPN (5.8)\\CHN (5.1)
    \end{tabular}  &  
    \begin{tabular}{l}
    Asset Management and Custody Banks (11.1)
    \end{tabular} &
    \begin{tabular}{l}
    IN (59.2)\\TE (39.7)
    \end{tabular}
    \\
    \midrule
    3:    &  525 & 
    \begin{tabular}{l}
    USA (22.3)\\IND (9.2)\\CHN (5.9)\\CAN (5.9)
    \end{tabular}  &  
    \begin{tabular}{l}
    Asset Management and Custody Banks (11.8)
    \end{tabular} &
    \begin{tabular}{l}
    IN (57.3)\\TE (41.9)
    \end{tabular}
    \\
    \midrule
    12:    &  403 & 
    \begin{tabular}{l}
    USA (18.9)\\IND (10.4)\\CAN (7.7)\\AUS (5.7)
    \end{tabular}  &  
    \begin{tabular}{l}
    Asset Management and Custody Banks (10.5)
    \end{tabular} &
    \begin{tabular}{l}
    IN (54.8)\\TE (44.2)
    \end{tabular}
    \\
    \midrule
    16:    &  364 & 
    \begin{tabular}{l}
    USA (24.4)\\CAN (11.3)\\GBR (8.7)\\IND (8.7)
    \end{tabular}  &  
    \begin{tabular}{l}
    Asset Management and Custody Banks (10.2)
    \end{tabular} &
    \begin{tabular}{l}
    IN (57.5)\\TE (41.2)
    \end{tabular}
    \\
    \bottomrule
    \end{tabular}
    \label{tab:com_OW}
%    \begin{itemize}
%    \color{red}{
%    \item The percentages for Primary Industry was wrong. I corrected them.}
%\end{itemize}
\end{table}
%+++++++++++++++++++++++++++++++++

%+++++++++++++++++++++++++++++++++
\begin{table}[thbp]
    \caption{Characteristics of the largest community at the 1st level for the ownership layer and the three largest communities at the 2nd level. The figures in brackets indicate the fraction of attribution, and we have only listed the attributions with a share above 5\%.}
    \vspace{3mm}
    \centering
    \begin{tabular}{lllll}
    \toprule
    \multicolumn{5}{l}{$G_3$: Ownership}  \\
    Index &  Size & Country & Primary Industry &  Bowtie \\
    \midrule
    1:1:    &  8,860 & 
    \begin{tabular}{l}
    USA (22.4)\\IND (8.8)\\CAN (6.2)\\CHN (5.8)\\GBR (5.4)\\JPN (5.1)
    \end{tabular}  &  
    \begin{tabular}{l}
    Asset Management and Custody Banks (10.8)
    \end{tabular} &
    \begin{tabular}{l}
    IN (56.2)\\TE (42.8)
    \end{tabular}
    \\
    \midrule
    1:2:    &  2,761 & 
    \begin{tabular}{l}
    USA (21.2)\\IND (9.1)\\CAN (6.7)\\AUS (6.1)\\CHN (5.1)\\GBR (5.1)
    \end{tabular}  &  
    \begin{tabular}{l}
    Asset Management and Custody Banks (10.0)
    \end{tabular} &
    \begin{tabular}{l}
    IN (58.1)\\TE (40.9)
    \end{tabular}
    \\
    \midrule
    1:5:    &  2,674 & 
    \begin{tabular}{l}
    USA (20.6)\\IND (9.1)\\CAN (7.6)\\GBR (5.5)\\CHN (5.4)\\AUS (5.3)\\JPN (5.1)
    \end{tabular}  &  
    \begin{tabular}{l}
    Asset Management and Custody Banks (9.4)
    \end{tabular} &
    \begin{tabular}{l}
    IN (55.3)\\TE (43.5)
    \end{tabular}
    \\
    \bottomrule
    \end{tabular}
    \label{tab:com_OW_2}
%    \begin{itemize}
%    \color{red}{
%    \item The percentages for Primary Industry was wrong. I corrected them.}
%\end{itemize}
\end{table}
%+++++++++++++++++++++++++++++++++

%=================================
\subsection*{Multilayer Analysis}
In the previous subsection, we described the characteristics of the isolated layers. 
To improve our understanding of the knowledge flows of drug pipelines, we investigate the relation between layers.

%=================
\paragraph{Interlayer Degree Correlations} 
The degrees of the same node in different layers can be correlated. For example, a node, that is, a hub, in the knowledge flows of drug pipelines is likely to be a hub in the supply-chain or ownership network as well. 
Based on the closed innovation in the pharmaceutical industry, how the pharmaceutical firms raise funds for the R\&D expenses against drug pipelines is seen as a significant problem. 
Hence, we can assume that companies have several drug pipelines. 
From this perspective, we compute the interlayer degree correlations between the knowledge flow layer and other layers.
Note that the edges in the knowledge flow layer $E_1 ^{p}$ have edge attributions with respect to the status of drug pipelines, as $p\in\{$Discovery, Phase I--III Clinical, Pre-registration, Registered, Launched$\}$, and only the ``Launched" status is cumulative.
Here we use the drug pipeline with launched $p=\{\text{Launched}\}$ because the other edges cannot tell us when the knowledge of drug pipelines flows.
The interpretation of edges in the knowledge flow layer we focus on is as follows:
\begin{itemize}
    \item Self-loops at the launched status, $\ell_i^{[1,p]}$ at $p=\{\text{Launched}\}$,\\
    corresponds to the number of drug pipelines the company $i$ discovered as the licenser and has launched the drugs.
    \item In-degree at the launched status removed self-loops, $m_{\text{in},i}^{[1,p]}$ at $p=\{\text{Launched}\}$,\\
    corresponds to the number of launched drug pipelines the company $i$ owns as the licensee.
\end{itemize}

\begin{figure}[thbp]
 \begin{minipage}{0.49\hsize}
  \begin{center}
   \includegraphics[width=2.5in]{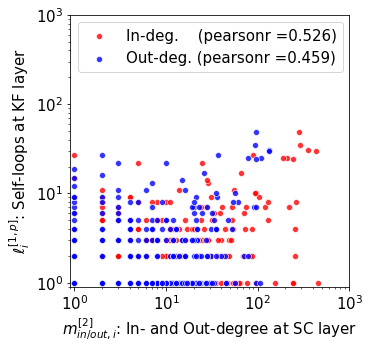}
  \end{center}
 \end{minipage}
 \begin{minipage}{0.49\hsize}
 \begin{center}
  \includegraphics[width=2.5in]{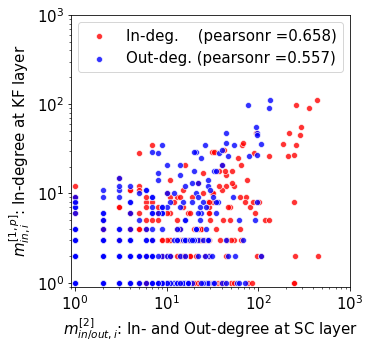}
 \end{center}
 \end{minipage}
 \begin{minipage}{0.49\hsize}
  \begin{center}
   \includegraphics[width=2.5in]{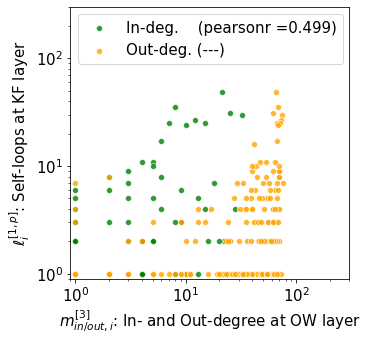}
  \end{center}
 \end{minipage}
 \begin{minipage}{0.49\hsize}
  \begin{center}
   \includegraphics[width=2.5in]{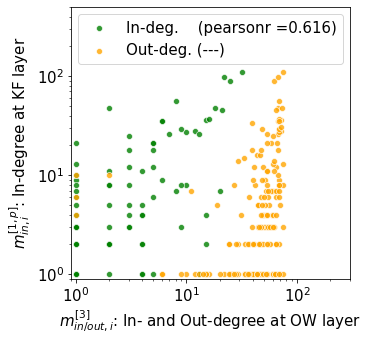}
  \end{center}
 \end{minipage}
 \caption{Interlayer degree correlations. Upper (lower) two figures illustrate the relations between the numbers of drug pipelines and the number of supplier-customer (ownership) links. The left (right) vertical axis corresponds to the number of  self-loops (in-degree) at the launched state in knowledge flow layer.}
 \label{fig:deg_cor}
\end{figure}

We show the relations between the number of drug pipelines and the number of supplier-customer (ownership) links in the upper (lower) part of Figure~\ref{fig:deg_cor}, using the Pearson correlation coefficients. 
There are upper limits of approximately 100 for the out-degree at the ownership layer, because of the threshold on data collection.
As anticipated, the companies having larger $m_{\text{in},i}^{[1,p]}$ at $p=\{\text{Launched}\}$ have a larger number of supplier-customer or ownership relations, and we found that there are weak but positive interlayer degree correlations.
However, there are also companies with a few suppliers and customers but a large $\ell_i^{[1,p]}$ at $p=\{\text{Launched}\}$.
Therefore, this result imply that the closed innovation in the pharmaceutical industry works well for large companies to keep the R\&D expenses and to sell them, but it is not necessary to guarantee the discovery of drug seeds.

%+++++++++++++++++++++++++++++++++
%+++++++++++++++++++++++++++++++++

%=================
\paragraph{Node and Edge Overlap} 
The primary purpose of our research is to understand the emergence of propagation and the localization of knowledge in the pharmaceutical industry, based on the supply-customer or ownership relation. 
The single-layer analysis for the knowledge flows of the drug pipeline demonstrates that approximately half of the drug pipelines are discovered and launched by individual countries or companies, and a comparison of the macroscopic structure suggests the similarity between the knowledge flow and supply chain in the pharmaceutical industry.
Furthermore, a hub in the knowledge flow network based on the drug pipeline tends to be a hub in the supply-chain or ownership layer as well. 
These results are consistent with a general understanding of how the localization of drug pipelines can be maintained under the condition of being costly, i.e., only larger firms, such as public companies having individual markets, can hold the drug pipelines. 
To improve our understanding regarding the propagation of knowledge between the pharmaceutical companies, we focus on the edge-level similarity characteristics of our MLN.

Although the dynamics of knowledge in drug pipeline flows between companies remain unclear, we can raise a possible hypothesis for the edge-level similarity. 
When drug pipelines are developing in an extremely closed situation, such that the pharmaceutical companies do not have business with the same industry, the edge-level similarities to the supply chain ceases to appear. 
However, the alliance by the pharmaceutical companies, i.e., open innovation in the pharmaceutical industry, has a potential to share not only markets but also drug pipelines, which is assumed to appear as the edge-level similarity to the supply-chain layer. 
Moreover, if the drug pipeline tends to be transferred to the owner of its licensor, we might observe the edge-level similarities to the ownership network.
Although the knowledge of the pharmaceutical industry may flow along with the flow of control, it is difficult to observe it in our MLN when the M\&A causes it. To confirm the above hypothesis, we compute the overlap of nodes and edges as 
\begin{equation}
    O(X_{\alpha},X_{\beta})=\frac{\left|X_{\alpha}\cap X_{\beta}\right|}{\left|X_{\alpha}\cup X_{\beta}\right|}~,
\end{equation}
where $X_{\alpha}$ is a set of nodes/edges at the $\alpha$-th layer.
We measure the overlap by the fraction of nodes/edges that appear in both layers over the aggregate number of nodes/edges of two layers.
We ignore the multiple edges and self-loops in the knowledge flow layer.  

\begin{figure}[thbp]
    \centering
    \includegraphics[width=4.5in]{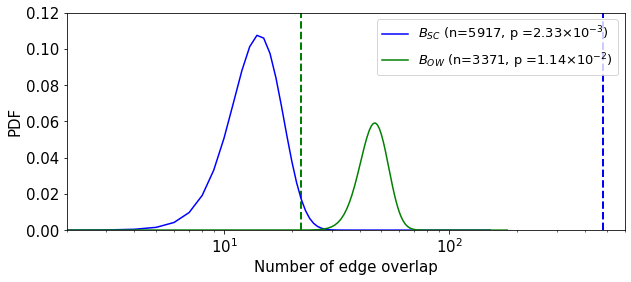}
    \caption{Comparison with binomial distribution. The solid lines represent binomial distributions $\mathrm{B}(n,p_{\alpha})$, where $p_{\alpha}$ is the conditional probability that we can choose the overlapped edge between knowledge flow layer and the $\alpha$-th layer from $n$ edges. The dashed lines represent actual values listed in Tabel~\ref{tab:overlap_ve}.}
    \label{fig:overlap}
\end{figure}

Table~\ref{tab:overlap_ve} shows the node and edge overlap between the two. 
We demonstrate the number of nodes and edges to compute the overlap values. 
The figures in brackets indicate the values computed for the GWCC for each layer. 
It is clear that there is a lower proportion of node/edge overlap present at the ownership layer, which is more remarkable as compared to the GWCC case. 
To check the statistical significance, we compute the probabilities (p-values) when the expected number of overlapped edges is larger than the observed value, using a statistical test.
%###########
Here, the null hypothesis is that the we have no edge overlap between the two layers. 
%###########
First, we can assume that the probability of generating the $\alpha$-th layer having the $x$ overlapping edges obeys the binomial distribution, $x\sim\text{B}(n,p(\beta|\alpha))$, where $n=\left|E_{\alpha}\cup E_{\beta}\right|$.
We define the conditional probability that the edge connects between two nodes overlapping knowledge flow layer ($\beta=1$) and the $\alpha$-th layer as 
\begin{equation}
     p(\beta=1|\alpha)=p_{1}\times p_{\alpha}=1\times\frac{\left<\bar{k}^{[\alpha]}\right>}{\left|V_{\alpha}\right|}~,
\end{equation}
where $\left<\bar{k}^{[\alpha]}\right>$ is the half value of the averaged total degree of the $\alpha$-th layer and we assume that the knowledge flow layer ($\beta=1$) is independent of the $\alpha$-th layer, $p_{1}=1$.
Now, the conditional probabilities $p(1|\alpha)$ are given as
\begin{equation}
     p(1|\alpha) = \begin{cases}
    2.33 \times 10^{-3} & (\alpha=2) ~ \\
    1.14 \times 10^{-2} & (\alpha=3) ~
  \end{cases}~
  \mathrm{where}~
  \left<\bar{k}^{[\alpha]}\right> = \begin{cases}
    2.68 & (\alpha=2) ~ \\
    1.01 & (\alpha=3) ~
  \end{cases}~,
\end{equation}
for the GWCC of supply chain $(\alpha=2)$ and ownership $(\alpha=3)$ network. 
Figure~\ref{fig:overlap} shows the comparisons of the actual number of edge overlaps with the above binomial distributions. 
%###########
%Consequently, the probability is almost zero, where the edge overlap between the two layers is equal %to the expected value obtained from the statistical test with the above binomial distribution. 
%This indicates that the edge overlaps between the knowledge flow layer and supply chain (ownership) %are significantly higher (lower).
%###########
Consequently, the p-value shows a value approximately equal to zero (one) for the supply-chain (ownership) layer. 
Therefore, the null hypothesis is rejected (adopted) for the supply-chain (ownership) layer. This indicates that the observed edge overlaps between the knowledge flow and the supply-chain (ownership) layers is (is not) statistically significant.
%###########

Thus, the knowledge flows in the pharmaceutical industry are related to the flow of products associated with the supply chain, rather than the dependency based on the ownership network. 
This suggests that the pharmaceutical companies use open innovation to share the drug pipeline and spread the business market as the supply chain.
%%%%%%%%%%%5
Although we do not use the ownership data including M\&A, the relation between the knowledge flow of the drug pipeline and ownership network is suggested to be rare. 
%%%%%%%%%%%5
The future work, therefore, should include a comparison with the traditional M\&A strategy of pharmaceutical companies in terms of their sustainable growth.

%+++++++++++++++++++++++++++++++++
\begin{table}[thbp]
\caption{Overlap of nodes and edges between the layers, as measured by the fraction of nodes/edges which appear in both layers over the aggregate number of nodes/edges of two layers. The multiple edges in the knowledge flow layer are ignored. The figures between brackets indicate the overlap/number of nodes/edges for GWCC of each layer.}
    \vspace{3mm}
    \centering
    \begin{tabular}{l|ccc}
    %\toprule
    {} &  Knowledge Flow &  Supply Chain &  Ownership \\
    %\midrule
    \midrule
    Knowledge Flow &           1.000 (1.000) &         {} &        {} \\
    Supply Chain   &           0.821 (0.730) &         1.000 (1.000) &        {} \\
    Ownership      &           0.342 (0.042) &         0.366 (0.057) &        1.000 (1.000)\\
    %\bottomrule
    \end{tabular}
    \\
    \vspace{3mm}
    (a) Node overlap between two layers \\
    \vspace{3mm}
    
    \begin{tabular}{l|ccc}
    %\toprule
    {} &  Knowledge Flow &  Supply Chain &  Ownership \\
    %\midrule
    \midrule
    Knowledge Flow &           1.000 (1.000) &         {} &        {} \\
    Supply Chain   &           0.067 (0.081) &         1.000 (1.000) &        {} \\
    Ownership      &           0.008 (0.007) &         0.015 (0.007) &        1.000 (1.000) \\
    %\bottomrule
    \end{tabular}
    \\
    \vspace{3mm}
    (b) Edge overlap between two layers \\
    \vspace{3mm}
    \begin{tabular}{l|ccc}
    {} &  Knowledge Flow &  Supply Chain &  Ownership \\
    %\midrule
    \midrule
    Knowledge Flow &        2,379 (1,574) &       {} &   {} \\
    Supply Chain   &        1,954 (1,149) &  1,954 (1,149)&   {} \\
    Ownership      &          813   (67) &      741 (66) &  813 (74)\\
    \end{tabular}
    \vspace{3mm}
    \\
    (c) Number of node overlap between two layers $\left|V_{\alpha}\cap V_{\beta}\right|$\\
    \vspace{3mm}
    \begin{tabular}{l|ccc}
    {} &  Knowledge Flow &  Supply Chain &  Ownership \\
    \midrule
    Knowledge Flow &            4,686 (3,318) &      {} &    {} \\
    Supply Chain   &             490 (480)  &    3,115 (3,079) &     {} \\
    Ownership      &              37 (22) &       47 (23) &    153 (75)\\
    \end{tabular}
    \vspace{3mm}
    \\
    (d) Number of edge overlap between two layers $\left|E_{\alpha}\cap E_{\beta}\right|$\\
    \label{tab:overlap_ve}
\end{table}
%+++++++++++++++++++++++++++++++++

%+++++++++++++++++++++++++++++++++
%+++++++++++++++++++++++++++++++++

%=========================================
\section{Conclusion}% Ikeda & Jibu
\label{sec:conclusion}
Open innovation is a lucrative field within industries relying on innovation.
Particularly, the pharmaceutical industry is a representation of knowledge-intensive industries, and open innovation is expected to play a significant role in boosting pharmaceutical collaborations and reducing the cost of long-term R\&D process or the M\&A in closed innovation.

Consequently, the development of a new drug and trial requires high investment; however, the success is likely to be a low possibility. 
Private companies, such as small and medium enterprises and ventures, end up funding innovation due to high investments. 
Moreover, public companies, such as big pharmaceutical companies, are always looking for new available targets. 
Thus, to boost pharmaceutical innovation, based on globalized and IT innovation, it is crucial to understand the dynamics of innovation in terms of various interactions of the firm-level economy.

To reveal the actual situation of pharmaceutical innovation, we proposed an MLN framework constructed with the drug pipeline, global supply chain, and ownership data. 
Furthermore, we investigated the features of the knowledge flows between the licenser and licensee, defined by the drug pipeline data, in terms of the proposed framework. 
Several seeds of the drug pipeline have been discovered in the USA and provided to the other countries. 
Larger firms, such as public companies having an individual market in the supply chain, can hold a significant number of drug pipelines as a licensee, which form cross-borders among pharmaceutical knowledge flows. 
The macroscopic structure of pharmaceutical knowledge flows provides us knowledge for features of each country. 
Since the firms in USA provide drug pipeline seeds to the whole system and those in Japan do not have many drug pipelines but deal with the second-largest launched drug pipeline, the US and JP firms appear on the upstream and the most substantial circular flow of knowledge flow network, respectively. 
Moreover, the tendency of closed innovation in CHN is seen not only as the self-sufficiency rate of drug pipelines but also as a closed community of pharmaceutical firms in CHN in the global supply-chain network.

The similarity between the knowledge flows of drug pipelines and supply-chain networks is also observed at the edge level, i.e., the licenser (licensee) of drug pipelines tends to be the supplier (customer) too, and a hub of knowledge flows of the drug pipelines tends to be their hub in the supply chain. 
Our results demonstrate a strong connection between an open innovation in the pharmaceutical industry and the firms’ activities in terms of the supply chain. In general, the pharmaceutical industries promote drug discovery and a clinical trial using artificial intelligence (AI) methods.  
For example, large pharmaceutical companies, such as Pfizer and Novartis, deployed AI systems for drug discovery and clinical trials by IBM \cite{top10ph, fleming2018artificial}. 
This collaboration suggests an increase in the needs for the relation between drug pipelines and supply chain in other industries. 
Based on the analysis, therefore, our findings agree with the reported situation of pharmaceutical innovation.

Finally, an essential question for future studies is to understand how some operations, such as policy-making, can boost pharmaceutical innovation. 
Thus, our study provides a framework for future studies to assess the features of pharmaceutical innovation in terms of various firm-level activities using MLN representation. Although our results are inspiring, more investigations are required for further improvements. 
It is possible to test the traditional M\&A strategy of pharmaceutical companies based on their sustainable growth, by combining the historical M\&A data with our networks. Moreover, our approach can investigate how the risk, patent cliff in the pharmaceutical industry will spread to the other sectors in the future.
We hope that future studies can contribute to reducing an unnecessary risk. 

%===========================================================
\section*{Acknowledgements}
We are grateful to Y. Fujiwara, H. Aoyama, H. Iyetomi, and H. Yoshikawa for their insightful comments and encouragement. 
The present study was supported by the Ministry of Education, Science, Sports, and Culture, Grants-in-Aid for Scientific Research (B), Grant No. 17KT0034 (2017-2019), and by MEXT as Exploratory Challenge s on Post-K computer (Studies of Multi-level Spatiotemporal Simulation of Socioeconomic Phenomena).

%===========================================================
\bibliographystyle{unsrt}  
\bibliography{references}

\end{document}